\documentclass{article}

\usepackage[T1]{fontenc}
\usepackage[utf8]{inputenc}
\usepackage[english]{babel}

\usepackage{amssymb}
\usepackage{amsmath}
\usepackage{amsthm}
\usepackage{cancel}
\usepackage{mathdots}

\usepackage{mathtools}

\usepackage{enumerate}

\def\proof{\begin{center} {\bf Proof:} \end{center}\vspace{0.5pt}}
\def\QEDclosed{\mbox{\rule[0pt]{1.3ex}{1.3ex}}} 
\def\QED{\QEDclosed} 
\def\endproof{\hspace*{\fill}~\QED\par\endtrivlist\unskip}

\newtheorem{thm}{Theorem}[]

\newtheorem{defn}[]{Definition}
\theoremstyle{definition}
\newtheorem*{rmk}{Remark}

\usepackage{palatino}
\begin{document}
\title{Degenerate first-order Hamiltonian operators of hydrodynamic type in 2D}
\author{Andrea Savoldi}
    \date{}
    \maketitle
    \vspace{-7mm}
\begin{center}
Department of Mathematical Sciences, Loughborough University \\
Leicestershire LE11 3TU, Loughborough, United Kingdom \\
e-mail: \\[1ex]
\texttt{A.Savoldi@lboro.ac.uk}\\
\end{center}

\bigskip

\begin{abstract}
First-order Hamiltonian operators of hydrodynamic type were introduced by Drubrovin and Novikov in 1983. In 2D, they are generated by a pair of contravariant metrics $g$, $\tilde g$ and a pair of differential-geometric objects $b$, $\tilde{b}$. If the determinant of the pencil $g+\lambda \tilde g$ vanishes for all $\lambda$, the operator is called degenerate. In this paper we provide a complete classification of degenerate two- and three-component Hamiltonian operators.
Moreover, we study the integrability, by the method of hydrodynamic reductions, of 2+1 Hamiltonian systems arising from the structures we classified.

\bigskip


\bigskip

\end{abstract}



\section{Introduction}
The theory of first-order Hamiltonian operator of differential-geometric type has been developed in the last three decades by several authors, starting from the pioneering work of Dubrovin and Novikov \cite{DN}. In one-dimensional case, these structures are given by
\begin{equation}\label{Ham1}
P^{ij}=g^{ij}({\bf u}) \frac{d}{dx} + b^{ij}_k ({\bf u}) u^k_x,
\end{equation}
where ${\bf u}=(u^1,\ldots, u^n)$ are local coordinates depending on $x$, $i,j,k=1,\ldots,n$, and $u^k_x=\frac{d u^k}{d x}$.
Dubrovin and Novikov proved that in the non-degenerate situation, namely $\det(g^{ij})\neq0$, \eqref{Ham1} defines a Poisson bracket through
$$
\{ F,G \} =\int \frac{\delta F}{\delta u^i} P^{ij} \frac{\delta G}{\delta u^j},
$$
if and only if $g^{ij}$ is a flat pseudo-Riemannian metric and the coefficients $\Gamma^{i}_{jk}=-g_{jm} b^{mi}_k$ are the Levi-Civita connection of the metric $g_{ij}$ (where $g^{i m}g_{mj}=\delta^{i}_{j}$). Thus, in flat coordinates, any non-degenerate Hamiltonian operator \eqref{Ham1} assumes constant form.
In the case where the metric $g$ is degenerate, that is, $\det(g^{ij})=0$, this result does not hold. Grinberg \cite{Grinberg} and later Bogoyavlenskij \cite{B1,B2} firstly investigated this case, and recently we provided a complete list of two- and three-component Poisson structures with degenerate metric \cite{Sav}.

First-order Hamiltonian operators of differential-geometric type naturally arise in the study of quasilinear systems (systems of hydrodynamic type). In 1+1 dimensions they are given by
$$
u^i_t+V^i_j({\bf u}) u^j_x=0.
$$
Such systems are called Hamiltonian if they can be written in the form
$$
u^i_t+P^{ij} \delta_j h=0,
$$
where $\delta_j=\delta / \delta u^j$ is the variational derivative, $h=h({\bf u})$ is the Hamiltonian density, and $P^{ij}$ is a Hamiltonian operator of hydrodynamic type \eqref{Ham1}.
It was conjectured by Novikov that a combination of the Hamiltonian property with the diagonalizability of the matrix $V^i_j$ implies the integrability. This conjecture was proved by Tsarev in \cite{Tsarev}, who established the linearizability of such systems by the generalised hodograph transform.

A generalisation of hydrodynamic type systems in 2+1 dimensions is given by
\begin{equation}\label{sys2+1_introduction}
{\bf u}_t + A({\bf u}) {\bf u}_x + B({\bf u}) {\bf u}_y=0,
\end{equation}
where ${\bf u}=(u^1,\ldots, u^n)$, $u^i=u^i(t,x,y)$ and $A, B$ are $n\times n$ matrices.
Systems of this type describe many physical phenomena. In particular, important examples occur in gas dynamics, shallow water theory, combustion theory, nonlinear elasticity, magneto-fluid dynamics, etc.
A system \eqref{sys2+1_introduction} is called Hamiltonian if it can be written in the form
$
u^i_t+P^{ij} \delta_j h=0,
$
where $P^{ij}$ is a 2D first-order Hamiltonian operator of differential-geometric type, namely
\begin{equation}\label{Ham2D_introduction}
P^{ij}= g^{ij}({\bf u}) \frac{d}{dx} + b^{ij}_k ({\bf u}) u^k_{x} +\tilde g^{ij}({\bf u}) \frac{d}{dy} + \tilde b^{ij}_k ({\bf u}) u^k_{y}.
\end{equation}

In 2+1 dimensions, a quasilinear system is said to be integrable if it can be decoupled in infinitely many ways into a pair of compatible $m$-component one-dimensional systems in Riemann invariants \cite{FK}. Ferapontov and Khusnutdinova proved that the requirement of the existence of sufficiently many $m$-component reductions provides an effective classification criterion. This method of hydrodynamic reductions, which is a natural analogue of the generalised hodograph transform in higher dimensions, leads to finite-dimensional moduli spaces of integrable Hamiltonians.

The purpose of this paper is two-fold. Starting from the classification of degenerate brackets in 1D, we want to describe degenerate Hamiltonian operators of hydrodynamic type in 2D, that is, operators of the form \eqref{Ham2D_introduction} such that $\det(g+\lambda \tilde g)=0$ holds $\forall \ \lambda$ (precise definition follows). Our analysis leads to a complete classification of two- and three-component degenerate structures (Section \ref{sect_deg}). Secondly, we study the integrability, by the method of hydrodynamic reductions, of Hamiltonian systems arising from three-component structures we classified (Section \ref{sect_sys}).

\section{Degenerate Hamiltonian operators in 2D}\label{sect_deg}

The problem of classification of multidimensional Hamiltonian operators was proposed by Dubrovin and Novikov in \cite{DN1}, and thoroughly investigated by Mokhov \cite{Mokhov1,Mokhov2}. Some results in the classification of 2D non-degenerate Hamiltonian operators were recently obtained in our paper \cite{FLS}.

A first-order multidimensional Hamiltonian operator of differential-gemetric type (Dubrovin-Novikov type) is defined by
\begin{equation}\label{HamMD}
P^{ij}=\sum_{\alpha=1}^{d} g^{ij\alpha}({\bf u}) \frac{d}{dx^{\alpha}} + b^{ij\alpha}_k ({\bf u}) u^k_{x^{\alpha}},
\end{equation}
where ${\bf u} = (u^1, \ldots, u^n)$ are local coordinates on a certain smooth $n$-dimensional manifold $M$ or a domain of $\mathbb{R}^n$, and
${\bf x} = (x^1, \ldots, x^d)$ are independent variables.

As in the one-dimensional case, the condition of skew-symmetry and the Jacobi identity for a Hamiltonian operator of the form \eqref{HamMD} impose very severe restrictions on the coefficients $g^{ij\alpha} ({\bf u})$ and $b^{ij\alpha}_k ({\bf u})$. In particular, Mokhov proved the following general statement:
\begin{thm}[\cite{Mokhov3}]\label{Mok}
Operator of the form \eqref{HamMD} is a Hamiltonian operator, i.e. it is skew-symmetric and satisfies the Jacobi identity, if and only if the following relations for the coefficients of the operator are fulfilled:
\begin{subequations}\label{mok_all}
\begin{equation}
g^{ij\alpha}=g^{ji\alpha},
\label{a1}
\end{equation}
\begin{equation}
\frac{\partial g^{ij\alpha}}{\partial u^k}=b^{ij\alpha}_k+b^{ji\alpha}_k,
\label{a2}
\end{equation}
\begin{equation}
\sum_{(\alpha,\beta)}\left (g^{si\alpha}b^{jr\beta}_s
-g^{sj\beta}b^{ir\alpha}_s\right )=0,
\label{a3}
\end{equation}
\begin{equation}
\sum_{(i,j,r)}\left (g^{si\alpha}b^{jr\beta}_s
-g^{sj\beta}b^{ir\alpha}_s\right )=0,
\label{a4}
\end{equation}
\begin{equation}
\sum_{(\alpha,\beta)}\left [g^{si\alpha}
\left (\frac{\partial b^{jr\beta}_s}{\partial u^q}
-\frac{\partial b^{jr\beta}_q}{\partial u^s}\right )
+b^{ij\alpha}_sb^{sr\beta}_q-b^{ir\alpha}_sb^{sj\beta}_q\right ]=0,
\label{a5}
\end{equation}
\begin{equation}
g^{si\beta}\frac{\partial b^{jr\alpha}_q}{\partial u^s}
-b^{ij\beta}_sb^{sr\alpha}_q-b^{ir\beta}_sb^{js\alpha}_q =
g^{sj\alpha}\frac{\partial b^{ir\beta}_q}{\partial u^s}
-b^{ji\alpha}_sb^{sr\beta}_q-b^{is\beta}_qb^{jr\alpha}_s,
\label{a6}
\end{equation}
\begin{gather}
\frac\partial{\partial u^k}
\left [g^{si\alpha}\left (\frac{\partial b^{jr\beta}_s}{\partial u^q}
-\frac{\partial b^{jr\beta}_q}{\partial u^s}\right )
+b^{ij\alpha}_sb^{sr\beta}_q-b^{ir\alpha}_sb^{sj\beta}_q\right ] 
+\sum_{(i,j,r)}\left [b^{si\beta}_q
\left (\frac{\partial b^{jr\alpha}_k}{\partial u^s}
-\frac{\partial b^{jr\alpha}_s}{\partial u^k}\right )\right ]\nonumber\\
+\frac\partial{\partial u^q}
\left [g^{si\beta}\left (\frac{\partial b^{jr\alpha}_s}{\partial u^k}
-\frac{\partial b^{jr\alpha}_k}{\partial u^s}\right )
+b^{ij\beta}_sb^{sr\alpha}_k-b^{ir\beta}_sb^{sj\alpha}_k\right ]
+\sum_{(i,j,r)}\left [b^{si\alpha}_k
\left (\frac{\partial b^{jr\beta}_q}{\partial u^s}
-\frac{\partial b^{jr\beta}_s}{\partial u^q}\right )\right ]=0.
\label{a7}
\end{gather}
\end{subequations}
\end{thm}
Relations (\ref{a1}) and (\ref{a2}) are equivalent to the skew-symmetry of the bivector \eqref{HamMD}, and relations (\ref{a3}){--}(\ref{a7})
are equivalent to the fulfilment of the Jacobi identity for a skew-symmetric bivector of the form \eqref{HamMD}.
The signs $\sum_{(\alpha, \beta)}$ and $\sum_{(i, j, k)}$ mean cyclic summation on the indicated indices.
Notice that for $n=1$, these conditions reduce to Grinberg's conditions \cite{Grinberg}.

In the one-dimensional case, Hamiltoinan operator \eqref{Ham1} is called \emph{degenerate} if $\det(g^{ij})=0$. In the multidimensional situation, we have the following
\begin{defn}
A $d$-dimensional operator of the form \eqref{HamMD} is said to be degenerate if  $\det \left(\sum_{\alpha=1}^d \lambda_{\alpha}g^{\alpha}\right)=0$ for any choice of $\lambda_1,\ldots, \lambda_d$, i.e. if there is no linear combination of the metrics $g^{\alpha}$ such that the determinant of this linear combination is non-zero.
\end{defn}
Let us point out that Theorem \ref{Mok} does not assume non-degeneracy of operators or additional conditions on the coefficients of \eqref{HamMD}. 

From Mokhov's conditions it immediately follows that each multidimensional Hamiltonian operator of the form \eqref{HamMD} is always the sum of one-dimensional Hamiltonian operators with respect to each of the independent variables $x^{\alpha}$, \cite{Mokhov3}.

Based on this result, and on the classification of one-dimensional degenerate Poisson structures of hydrodynamic type, we give a complete description of two- and three-component degenerate Hamiltonian operators for $d=2$, namely
\begin{equation}\label{Ham2D}
P^{ij}= g^{ij}({\bf u}) \frac{d}{dx} + b^{ij}_k ({\bf u}) u^k_{x} +\tilde g^{ij}({\bf u}) \frac{d}{dy} + \tilde b^{ij}_k ({\bf u}) u^k_{y}.
\end{equation}
For simplicity, let us label the $x$-part and the $y$-part of the Hamiltonian operator \eqref{Ham2D} respectively with $P_{(x)}$ and $P_{(y)}$ .

Hamiltonian structure of the form \eqref{Ham2D} is called \emph{trivial} if it is identically zero, or if it can be reduced to the form
\begin{equation}\label{multiply}
\tilde{g}^{ij}=\xi g^{ij}, \quad \tilde b^{ij}_k=\xi b^{ij}_k,
\end{equation}
for $\xi$ constant.
Notice that allowing linear change of the independent variables $x,y$, an operator satisfying \eqref{multiply} is essentially 1D.

\begin{rmk}
Let us remark that if a pair of Hamiltonian operators defines a 2D structure, by \eqref{mok_all} it easily follows that these two operators are compatible, and therefore they define a bi-Hamiltonian structure \cite{Mokhov2,Mokhov3}. Degenerate bi-Hamiltonian structures of hydrodynamic type were firstly investigated by Strachan \cite{Strachan1,Strachan}, revealing a nice relation with the theory of the analogous of Frobenius manifolds with degenerate metric.
\end{rmk}

\subsection{Classification}
The analysis of Mokhov's conditions \eqref{mok_all} is not straightforward. In order to study two- and three-component structures, we fix the pair $(g,b)$ given by the classification of 1D Hamiltonian operators \cite{Sav}. This classification can be summarised in the following two theorems.
\begin{thm}\label{thm2cmpt}
Any degenerate two-component Hamiltonian operator of Dubrovin-Novikov type in 1D can be brought, by a change of the dependent variables, to one of the following two canonical forms:
\begin{equation}\label{1D_2cmpt}
P=\begin{pmatrix}
d_x & 0\\
0 & 0
\end{pmatrix},
\quad
P=
\begin{pmatrix}
d_x & -\dfrac{u^2_x}{u^1}\\
\dfrac{u^2_x}{u^1} & 0
\end{pmatrix}.
\end{equation}
\end{thm}

\begin{thm}\label{thm3cmpts}
Any degenerate three-component Hamiltonian operator of Dubrovin-Novikov type in 1D can be brought, by a change of the dependent variables, to one of the following canonical forms:
\begin{itemize}
\item rank(g) = 0:
\end{itemize}
\begin{equation}\label{rank0}
P=\begin{pmatrix}
0 & u^3_x & 0\\
-u^3_x & 0 & 0\\
0 & 0 & 0
\end{pmatrix},
\end{equation}
\begin{itemize}
\item rank(g) = 1:
\end{itemize}
\begin{equation}\label{rank1}
P=\begin{pmatrix}
d_x & 0 & 0\\
0 & 0 & 0\\
0 & 0 & 0
\end{pmatrix},
\;
P=\begin{pmatrix}
d_x & u^3_x& 0\\
-u^3_x& 0 & 0\\
0& 0 & 0
\end{pmatrix},
\;
P=
\begin{pmatrix}
d_x & 0& -\frac{u^3_x}{u^1}\\
0& 0 & 0\\
\frac{u^3_x}{u^1}& 0 & 0
\end{pmatrix},
\;
P=
\begin{pmatrix}
d_x & -\frac{u^2_x}{u^1} & -\frac{u^3_x}{u^1}\\
\frac{u^2_x}{u^1} & 0 & 0\\
\frac{u^3_x}{u^1} & 0 & 0
\end{pmatrix},
\end{equation}
\begin{itemize}
\item rank(g)=2:
\end{itemize}
\begin{equation}\label{rank2}
\begin{array}{c}
P=\begin{pmatrix}
0 & d_x & 0\\
d_x & 0 & 0\\
0 & 0 & 0
\end{pmatrix},
P=\begin{pmatrix}
0 & d_x & -\frac{u^3_x}{u^2}\\
d_x & 0 & 0\\
\frac{u^3_x}{u^2} & 0 & 0
\end{pmatrix},
P=\begin{pmatrix}
0 & d_x & \frac{u^3_x}{u^3u^1-u^2} \\
d_x & 0 & \frac{-u^3 u^3_x}{u^3u^1-u^2} \\
\frac{-u^3_x}{u^3 u^1-u^2}& \frac{u^3 u^3_x}{u^3u^1-u^2} & 0
\end{pmatrix}.
\end{array}
\end{equation}
\end{thm}
Once we have fixed the pair $(g,b)$, solving \eqref{mok_all} we are able to find the pair $(\tilde{g}, \tilde{b})$. At this point, we look for canonical forms of 2D structures using transformations which preserve the form of the first structure given by $(g,b)$. As we will see, in some cases these transformations are not enough to eliminate all the functional parameters appearing in the 2D structure.

Let us agree on some notation: if a function depends only on one variable, we denote with $'$ the derivative with respect to that variable. Otherwise, if a function depends on more than one variable, say, $f=f(u^1,\ldots, u^n)$, then we use $\partial_i f= \frac{\partial f}{\partial u^i}$. In Section \ref{sect_sys}, for simplicity, the derivative with respect to $u^i$ will be denote as $f_{u^i}$.

\subsubsection{Two-component case}
Here we provide a full description of the two-component case.
\begin{thm}\label{thm_2D_2cmpt}
Any non-trivial degenerate two-component Hamiltonian operator of Dubrovin-Novikov type in 2D can be brought, by a change of the dependent variables, to the following form
\begin{equation}\label{2D_2cmpt_1}
P=\begin{pmatrix}
d_x+u^2 d_y +\frac{1}{2}u^2_y & -\epsilon\frac{u^2_x +u^2 u^2_y}{u^1}\\
\epsilon\frac{u^2_x +u^2 u^2_y}{u^1} & 0
\end{pmatrix},
\end{equation}
where $\epsilon$ can be either $0$ or $1$.
\end{thm}

\proof
First of all, the case  $g=\tilde g=0$ gives no non-trivial solutions.
In the case where the rank of the pencil $g^{ij}+\lambda \tilde g^{ij}$ is constantly equal to one, there exists a coordinates system $(u^1, u^2)$ where
$$
g^{ij}=
\begin{pmatrix}
1 &0\\
0&0
\end{pmatrix}, \quad
\tilde g^{ij}=
\begin{pmatrix}
f &0\\
0&0
\end{pmatrix},
$$
here $f=f(u^1,u^2)$ is some function. Let us fix the $P_{(x)}$ structure.

\medskip \noindent
{\bf Case} $\eqref{1D_2cmpt}_1$. If $b^{ij}_k$ are all identically zero, conditions \eqref{mok_all} imply
$$
f=f(u^2), \quad \tilde b^{11}_2=\frac{f'}{2},
$$
and all other $\tilde b^{ij}_k$ equal to zero. If $f=\xi$ is constant, than $\tilde g=\xi g$. Otherwise, using a transformation which preserves $P_{(x)}$, that is, a suitable change of coordinates of the form
$
u^1=v^1+\varphi^1(v^2),  \ u^2=\varphi^2(v^2),
$
we can easily reduce $f$ to $v^2$, obtaining \eqref{2D_2cmpt_1} with $\epsilon=0$.

\medskip \noindent
{\bf Case} $\eqref{1D_2cmpt}_2$. Suppose $b^{21}_2=-b^{12}_2=\frac{1}{u^1}$. Conditions \eqref{mok_all} imply
$$
f=f(u^2), \quad
\tilde b^{11}_2=\frac{f'}{2}, \quad \tilde b^{21}_2=-\tilde b^{12}_2=\frac{f}{u^1}.
$$
If $f=\xi$ is constant, than $\tilde g=\xi g$, $\tilde b=\xi b$. Otherwise, let us assume $f$ non-constant.
Transformations which preserve $P_{(x)}$ are given by $u^1=v^1, \ u^2=\varphi(v^2)$, then we can always choose $\varphi$ such that $f$ reduces to $v^2$ in the new coordinate system, obtaining \eqref{2D_2cmpt_1} with $\epsilon=1$.
\endproof

\subsubsection{Three-component case}
The analysis of the three-component situation is more complicated. Let us consider separately the cases with respect to the rank of the pencil $g_{\lambda} \! = \! g \! - \!\lambda \tilde g$. We point out that in some cases the group of transformations preserving the first structure $P_{(x)}$ is not sufficient to reduce the second structure $P_{(y)}$ to something simpler. Thus, we will just consider the more general structure given by the solution of Mokhov's conditions. The results can be stated as follows.

\begin{thm}\label{thm_2D_3cmpt_rank0}
$\mathrm{Rank}(g_{\lambda})=0$. Any non-trivial degenerate three-component Hamiltonian operator of Dubrovin-Novikov type in 2D can be brought, by a change of the dependent variables, to one of the following forms:
\begin{equation}\label{rank0_P}
P=
\begin{pmatrix}
0 & u^3_x +u^1 u^3_y  & 0\\
-u^3_x-u^1 u^3_y  & 0 & 0\\
0 & 0 & 0
\end{pmatrix},
\quad
P=
\begin{pmatrix}
0 & u^3_x +u^3 u^3_y & 0\\
-u^3_x - u^3 u^3_y & 0 & 0\\
0 & 0 & 0
\end{pmatrix}.
\end{equation}
\end{thm}
In this case, we do not need any linear change of the independent variables $x, y$.

\begin{thm}\label{thm_2D_3cmpt_rank1}
$\mathrm{Rank}(g_{\lambda})=1$. Any non-trivial degenerate three-component Hamiltonian operator of Dubrovin-Novikov type in 2D can be brought, by a change of the dependent variables and linear change of $x$ and $y$, to one of the following forms:
\begin{equation}\label{rank1_P_1}
P=
\begin{pmatrix}
d_x+\epsilon\left(u^2 d_y +\frac{u^2_y}{2}\right)& 0 & h u^2_y \\
0 & 0 & 0\\
- h u^2_y & 0 & 0
\end{pmatrix},
P=
\begin{pmatrix}
d_x+f d_y +\frac{\partial_2 f u^2_y +\partial_3 f u^3_y }{2}& 0 & -\frac{u^3_x-h u^2_y +f u^3_y}{u^1}\\
0 & 0 & 0\\
\frac{u^3_x-h u^2_y +f u^3_y}{u^1}& 0 & 0
\end{pmatrix},
\end{equation}
\begin{equation}\label{rank1_P_2}
P=
\begin{pmatrix}
d_x+f d_y +\frac{\partial_2 f u^2_y + \partial_3 f u^3_y}{2}& u^3_x +h u^3_y & 0\\
-u^3_x -h  u^3_y & 0 & 0\\
0 & 0 & 0
\end{pmatrix},
P=
\begin{pmatrix}
d_x+u^2 d_y +\frac{u^2_y}{2} & -\frac{u^2_x +u^2 u^2_y}{u^1} & -\frac{u^3_x+u^2 u^3_y}{u^1} \\
\frac{u^2_x +u^2 u^2_y}{u^1}  & 0 & 0\\
\frac{u^3_x +u^2 u^3_y}{u^1}  & 0 & 0
\end{pmatrix},
\end{equation}
where $f=f(u^2,u^3)$, $h=h(u^2,u^3)$ are arbitrary functions and $\epsilon$ can be either $0$ or $1$.
\end{thm}

\begin{thm}\label{thm_2D_3cmpt_rank2}
$\mathrm{Rank}(g_{\lambda})=2$. Any non-trivial degenerate three-component Hamiltonian operator of Dubrovin-Novikov type in 2D can be brought,  by a change of the dependent variables and linear change of $x$ and $y$, to one of the following forms:
\begin{equation}\label{rank2_P_1}
P=
\begin{pmatrix}
-2 u^1\ d_y -u^1_y & d_x+ u^2 \ d_y +2 u^2_y &  \epsilon u^3_y\\
d_x+ u^2 \ d_y -u^2_y& 0 & 0\\
-\epsilon u^3_y& 0 & 0
\end{pmatrix},
\quad
P=
\begin{pmatrix}
0 & d_x& d_y \\
d_x &0 & 0\\
d_y  & 0 & 0
\end{pmatrix},
\end{equation}
\begin{equation}\label{rank2_P_2}
P=
\begin{pmatrix}
p\ d_y +\frac{p' u^3_y}{2}&d_x + q \ d_y +\epsilon u^3_y & 0\\
d_x+ q\ d_y +(q'-\epsilon) u^3_y& r \ d_y +\frac{r' u^3_y}{2} & 0\\
0 & 0 & 0
\end{pmatrix},
\quad
P=
\begin{pmatrix}
d_y & d_x & -\frac{u^3_x}{u^2}\\
d_x & 0 & 0\\
\frac{u^3_x}{u^2} & 0 & 0
\end{pmatrix},
\end{equation}
\begin{equation}\label{rank2_P_3}
P=
\begin{pmatrix}
\epsilon d_y & d_x +u^3 \ d_y & -\frac{u^3_x+u^3 u^3_y}{u^2} \\
d_x +u^3 \ d_y + u^3_y& 0 & 0\\
\frac{u^3_x+u^3 u^3_y}{u^2}&0 &0
\end{pmatrix},
\quad
P=
\begin{pmatrix}
0 & d_x & -\frac{u^3_x-u^1_y}{u^2} \\
d_x & 0 & d_y \\
\frac{u^3_x-u^1_y}{u^2} & d_y & 0
\end{pmatrix},
\end{equation}
\begin{equation}\label{rank2_P_4}
P=
\begin{pmatrix}
0 & d_x& d_y -\frac{u^3_x-u^2_y }{u^2} \\
d_x  & 0 & 0\\
d_y+\frac{u^3_x-u^2_y}{u^2}&0 &0
\end{pmatrix},
\quad
P=
\begin{pmatrix}
u^1\ d_y +\frac{u^1_y}{2}& d_x -\frac{u^2}{2}  \ d_y - u^2_y & -\frac{u^3_x}{u^2} \\
d_x-\frac{ u^2}{2}  \ d_y +\frac{u^2_y}{2}& 0 & 0\\
\frac{u^3_x}{u^2} &0 &0
\end{pmatrix},
\end{equation}
\begin{equation}\label{rank2_P_5}
P=
\begin{pmatrix}
d_y &d_x- u^3  \,  d_y& \frac{u^3_x-2 u^3 u^3_y }{u^3 u^1-u^2}\\
 d_x - u^3 \,  d_y-u^3_y&  (u^3)^2  \, d_y+u^3 u^3_y & -\frac{u^3 u^3_x -2 (u^3)^2  u^3_y }{u^3 u^1-u^2}\\
-\frac{u^3_x -2 u^3 u^3_y }{u^3 u^1-u^2}&\frac{u^3 u^3_x -2 (u^3)^2 u^3_y }{u^3 u^1-u^2}&0
\end{pmatrix},
\end{equation}
\begin{equation}\label{rank2_P_6}
P=
\begin{pmatrix}
\frac{\kappa \, d_y}{u^3}-\frac{\kappa u^3_y}{(u^3)^2}& d_x- \kappa d_y +\frac{\kappa u^3_y}{2 u^3}& \frac{u^3_x -2 \kappa u^3_y }{u^3 u^1-u^2}\\
d_x- \kappa \,d_y-\frac{\kappa u^3_y}{2 u^3}&  \kappa u^3 \, d_y +\frac{u^3_y}{2} & -\frac{u^3 u^3_x-2 \kappa u^3 u^3_y }{u^3 u^1-u^2}\\
-\frac{u^3_x-2 \kappa u^3_y }{u^3 u^1-u^2}&\frac{u^3 u^3_x-2 \kappa u^3 u^3_y }{u^3 u^1-u^2}&0
\end{pmatrix},
\end{equation}
where $p,q,r$ are arbitrary functions on $u^3$, $\kappa$ is constant and $\epsilon$ can be either $0$ or $1$.
\end{thm}
The proof of these theorems can be found in the Appendix.

Let us point out that, after swapping the coordinates $u^1$, $u^2$, $\eqref{rank2_P_3}_2$ corresponds to the Hamiltonian operator for the 2D equations of gas dynamic (see, for instance, \cite{FK1}), namely
\begin{equation}\label{P_gas}
P^{ij}=
\begin{pmatrix}
0 &d_x & d_y \\
d_x & 0 & \frac{u^2_y-u^3_x}{u^1} \\
d_y & \frac{u^3_x-u^2_y}{u^1} & 0
\end{pmatrix},
\end{equation}
We will discuss it in Section \ref{sect_gas}.


\section{Hamiltonian systems of hydrodynamic type in 2+1 dimensions}\label{sect_sys}
In this section we discuss (2+1)-dimensional Hamiltonian systems of hydrodynamic type,
\begin{equation}\label{sys2+1}
{\bf u}_t + A({\bf u}) {\bf u}_x + B({\bf u}) {\bf u}_y=0,
\end{equation}
which are representable in the form ${\bf u}_t+P h_{\bf u}=0$, where $h({\bf u})$ is a Hamiltonian density and $P$ is a two-dimensional Hamiltonian operator of differential-geometric type \eqref{Ham2D}. As we recalled in the introduction, a (2+1)-dimensional quasilinear system is said to be integrable if it can be decoupled in infinitely many ways into a pair of compatible $m$-component one-dimensional systems in Riemann invariants. Let us briefly describe the method of hydrodynamic reduction introduced by Ferapontov and Khusnutdinova in \cite{FK}.

\subsection{The method of hydrodynamic reductions}
The method of hydrodynamic reductions is based on the existence of exact solutions of the (2+1)-dimensional system \eqref{sys2+1}. These solutions have the form ${\bf u}={\bf u}(R^1, ..., R^m)$, where the Riemann invariants ${\bf R}=(R^1, ..., R^m)$ solve a pair of commuting diagonal systems
\begin{equation}\label{riem_inv}
R^i_t=\lambda^i({\bf R})\ R^i_x, \quad R^i_y=\mu^i({\bf R})\ R^i_x.
\end{equation}
Let us point out that we do not impose any constraint on the number of Riemann invariants: $m$ is arbitrary.
Therefore, the (2+1)-dimensional system we are considering \eqref{sys2+1}, is decoupled into a pair of diagonal (1+1)-dimensional systems given by \eqref{riem_inv}. Usually, these solutions are known as nonlinear interactions of $m$ planar simple waves.

It turns out that the commutativity of the flows \eqref{riem_inv} is equivalent to the following constraints on the characteristic speeds $\lambda^i, \mu^i$ \cite{Tsarev}:
\begin{equation} \label{comm}
\frac{\partial_j\lambda^i}{\lambda^j-\lambda^i}=\frac{\partial_j\mu^i}{\mu^j-\mu^i}, ~~~ i\ne j, ~~~  \partial_j=\frac{\partial}{\partial {R^j}},
\end{equation}
(no summation). Imposing these restrictions, the general solution of systems \eqref{riem_inv} is given by the implicit generalised hodograph formula \cite{Tsarev}
\begin{equation}\label{hod}
v^i({\bf R})=x+\lambda^i({\bf R})\ t+\mu^i({\bf R}) \ y, \quad i=1, ..., m.
\end{equation}
Here the functions $v^i({\bf R})$ are characteristic speeds of the general flow commuting with \eqref{riem_inv}, namely, the general solution of the linear system
\begin{equation} \label{comm1}
\frac{\partial_jv^i}{v^j-v^i}=\frac{\partial_j\lambda^i}{\lambda^j-\lambda^i}=\frac{\partial_j\mu^i}{\mu^j-\mu^i}, \quad i\ne j.
\end{equation}
By straightforward computation, the substitution of ${\bf u}(R^1, ..., R^m)$ into \eqref{sys2+1}, using \eqref{riem_inv}, leads to
\begin{equation}\label{sys_21d}
(E + \lambda^i A+\mu^i B) \partial_i{\bf u}=0, \quad i=1, ..., m,
\end{equation}
where $E$ is the $n\times n$ identity matrix. This means that both $\lambda^i$ and $\mu^i$ have to satisfy the dispersion relation
\begin{equation} \label{dispersion}
\det (E + \lambda^i A+\mu^i B)=0.
\end{equation}
Furthermore, the construction of nonlinear interactions of $m$ planar simple waves can be summarised as follows.
First of all, we have to decoupled the initial (2+1)-dimensional system \eqref{sys2+1} into a pair of commuting flows \eqref{riem_inv}, by solving the equations  \eqref{comm}, \eqref{sys_21d} for ${\bf u}({\bf R}), \ \lambda^i({\bf R}), \ \mu^i({\bf R})$ as functions depending on the Riemann invariants $R^1, ..., R^m$. It is not difficult to see that for $m\geq 3$ the system given by these equations is overdetermined. Thus, in general this system does not posses solutions. However, if we are able to construct a particular reduction of the form \eqref{riem_inv}, the second step is quite straightforward: we have to solve the linear system given by \eqref{comm1} for the functions $v^i({\bf R})$, and then we can obtain $R^1, ..., R^m$ as functions of $t, x, y$ from the implicit  hodograph formula \eqref{hod}.

What can we say about the number of $m$-component reductions that a (2+1)-dimensional system \eqref{sys2+1} may admit?  
Analysing equations \eqref{comm} and \eqref{sys_21d}, one can prove that this number is parametrised, up to changes of variables of the form $R^i\to f^i(R^i)$, by $m$ arbitrary functions of a single variable. Remarkably, this number does not depend on $n$. This leads to the following definition.
\begin{defn}[\cite{FK}]
A (2+1)-dimensional quasilinear system is said to be integrable if it possesses $m$-component reductions of the form \eqref{riem_inv}  parametrised  by $m$ arbitrary functions of a single argument.
\end{defn}

\begin{rmk}
Looking at the structure of equations \eqref{comm} and \eqref{sys_21d}, one can see that their consistency conditions involve only triple of indices $i\ne j \ne k$. Moreover, all these conditions are completely symmetric in $i,j$ and $k$, and then it is enough to verify them setting, for instance, $i=1,\ j=2,\ k=3$. This means that the existence of non-trivial three-component reductions implies the existence of  $m$-component reductions for arbitrary $m$.
\end{rmk}

\begin{rmk}
We require that $\lambda^i$ and $\mu^i$ do not satisfy any linear relation, otherwise we would have no sufficiently many arbitrary functions of a single argument. Indeed, let us suppose that $\mu^i=a \lambda^i+b$. Condition \eqref{comm} reads $\partial_j a \lambda^i +\partial_jb=0$, which implies $a$ and $b$ constant. Thus, solutions of the system \eqref{riem_inv}, 
$$
R^i_t=\lambda^i R^i_x, \quad R^i_y=(a\lambda^i +b)R^i_x,
$$
are of the form $R^i=R^i(x+by,t+ay)$. These solutions correspond to travelling wave reduction, and they clearly do not contain enough arbitrary functions.
\end{rmk}

\subsection{Generalised two-dimensional gas dynamic equations}\label{sect_gas}
The equations of two-dimensional isentropic gas dynamics are of the form
\begin{equation} \label{syst_ex}
\rho_t + (\rho u )_x + (\rho v )_y =0, \quad
u_t+ u u_x + v u_y + \frac{p_x}{\rho}=0, \quad
v_t+ u v_x + v v_y + \frac{p_y}{\rho}=0,
\end{equation}
where $p=p(\rho)$ is the equation of state. In matrix form \eqref{sys2+1}, one has $\textbf{u}=(\rho, u, v)^t$ and
\begin{gather*}
A=
\left(\begin{array}{ccc}
u & \rho & 0 \\[6pt]
\dfrac{c^2}{\rho} & u & 0 \\[6pt]
0 & 0  & u
\end{array}\right),
\quad
B=
\left(\begin{array}{ccc}
v & 0 & \rho\\
0 & v & 0\\
\dfrac{c^2}{\rho} & 0 & v
\end{array}\right),
\end{gather*}
where $c^2=p^{\prime} (\rho)$ is the sound speed.
As demonstrated in \cite{SSY}, there exist potential flows describing nonlinear interaction of two sound waves which are locally parametrised by four arbitrary functions of a single argument.

The system \eqref{syst_ex} can be written in Hamiltonian form as ${\bf u}_t + P h_{\bf u}=0$, where the operator $P$ is given by \eqref{P_gas}, namely
$$
P^{ij}=
\begin{pmatrix}
0 & d_x & d_y \\
d_x & 0 & \frac{u_y-v_x}{\rho} \\
d_y & \frac{v_x-u_y}{\rho} & 0
\end{pmatrix},
$$
the Hamiltonian density $h$ is $h(\rho, u, v)=\frac{1}{2} \rho (u^2+v^2)+k(\rho)$, and the equation of state $p$ and the function $k$ are related by $p_\rho=\rho k_{\rho\rho}$.

Let us assume $h=h(\rho, u, v)$ generic, thus the system  ${\bf u}_t + P h_{\bf u}=0$ reads
\begin{equation}\label{sys_1}
\rho_t+(h_u)_x+(h_v)_y=0, \quad
u_t+(h_{\rho})_x+\frac{u_y-v_x}{\rho} h_v =0, \quad
v_t+(h_{\rho})_y+\frac{v_x-u_y}{\rho} h_u=0.
\end{equation}
Let us consider the Riemann invariants $R^1, \ldots, R^m$ solving
\begin{equation*}
R^i_x= \lambda^i({\bf R}) R^i_t, \quad R^i_y= \mu^i ({\bf R}) R^i_t, \quad i=1,\ldots, m.
\end{equation*}
By straightforward computation, the substitution $\rho=\rho({\bf R}), u=u({\bf R}), v=v({\bf R})$ into \eqref{sys_1} implies
\begin{gather}
(1+ \lambda^i  h_{\rho u} + \mu^i h_{\rho v})  \partial_i u + h_{\rho \rho} \lambda^i \partial_i \rho=0, \label{eq_partial_u}\\
(1+ \lambda^i  h_{\rho u} + \mu^i h_{\rho v})  \partial_i v + h_{\rho \rho} \mu^i \partial_i \rho=0, \label{eq_partial_v}\\ 
(1+ \lambda^i  h_{\rho u} + \mu^i h_{\rho v}) \partial_i \rho + (\lambda^i h_{uu} + \mu^i h_{uv}) \partial_i u +(\lambda^i h_{uv} + \mu^i h_{vv}) \partial_i v=0,
\end{gather}
here $i=1,\ldots, m, \; \partial_i=\frac{\partial}{\partial R^i}$.
Note that since $ \mu^i \partial_i u= \lambda^i \partial_i v$ (this easy follows from  \eqref{eq_partial_u} and  \eqref{eq_partial_v}, assuming $1+ \lambda^i  h_{\rho u} + \mu^i h_{\rho v}\neq0$), one has $ u_y=v_x$.
Thus, solutions are necessarily potential. Then, setting $u=\varphi_x$ and $v=\varphi_y$, our system \eqref{sys_1} reads
\begin{equation}\label{sys_1_new}
\rho_t +(h_u)_x+(h_v)_y=0, \quad
\varphi_{xt }+(h_{\rho})_x=0, \quad
\varphi_{yt} +(h_{\rho})_y=0.
\end{equation}
Both the last two equations give $\varphi_t+h_{\rho}=0$, so we finally have the following system
\begin{equation}\label{sys_1_2eq} 
\rho_t+(h_u)_x+(h_v)_y=0, \quad \varphi_t+h_{\rho}=0.
\end{equation}
If we consider the partial Legendre transform
\begin{equation} \label{legendre}
\tilde{\rho}= h_{\rho}, \quad
\tilde{u}=u, \quad
\tilde{v}=v, \quad
\tilde{h}=h - \rho h_{\rho},
\end{equation}
the derivatives respect the new variables are
\begin{equation}\label{eq_tilde}
\tilde{ h}_{\tilde{\rho}}=-\rho, \quad \tilde{h}_{\tilde{u}}= h_{u}, \quad \tilde{h}_{\tilde{v}}= h_{v},
\end{equation}
and we can rewrite the system \eqref{sys_1_2eq} in the form
$$
\big( \tilde{h}_{\tilde{\rho}}\big)_t+\big( \tilde{h}_{\tilde{u}}\big)_x +\big( \tilde{h}_{\tilde{v}}\big)_y=0,\\
\quad \varphi_t=\tilde{\rho}, \quad \varphi_x=\tilde{u}, \quad \varphi_y=\tilde{v}.
$$
The function $\tilde{h}$ depends only on $\varphi_x, \varphi_y, \varphi_t$ and thus we obtain three-dimensional Euler-Lagrange equations (setting $\tilde{h}=f$)
\begin{gather}
\left( f_{\varphi_x}\right)_x+\left( f_{\varphi_y}\right)_y+\left( f_{\varphi_t}\right)_t=0, \label{eq_lagrange}
\end{gather}
corresponding to Lagrangian densities of the form $f(\varphi_x, \varphi_y, \varphi_t)$. For example, the Lagrangian density $f=u_x^2+u_y^2-2e^{u_t}$ leads to the Boyer-Finley equation $u_{xx}+u_{yy}=e^{u_t}u_{tt}$ \cite{BF}.

In \cite{FKT} Ferapontov, Khusnutdinova and Tsarev derived a system of partial differential equations for the Lagrangian density $f(\varphi_x, \varphi_y, \varphi_t)$ which are  necessary and sufficient  for the integrability of the equation \eqref{eq_lagrange} by the method of hydrodynamic reductions (see also \cite{FO} for further details). Setting $a=\varphi_x, b=\varphi_y, c=\varphi_t$, these conditions  can be represented in a remarkable compact form:

\begin{thm}[\cite{FKT}]\label{thm_Fer}
For a  non-degenerate Lagrangian, the Euler-Lagrange equation \eqref{eq_lagrange} is integrable by the method of hydrodynamic reductions if and only if the density $f$ satisfies the relation
\begin{equation}
d^4f=d^3f\frac{dH}{H}+\frac{3}{H}\det(dM);
\label{fourth}
\end{equation}
here $d^3f$ and $d^4f$ are the symmetric differentials of $f$. The Hessian $H$ and the $4 \times 4$ matrix $M$ are defined as follows:
\begin{equation}
H=det
\left(\begin{array}{ccc}
f_{aa} & f_{ab} & f_{ac} \\
f_{ab} & f_{bb} & f_{bc} \\
f_{ac} & f_{bc} & f_{cc}
\end{array}
\right), ~~~
M=\left(\begin{array}{cccc}
0 & f_a & f_b & f_c \\
f_a & f_{aa} & f_{ab} & f_{ac} \\
f_b & f_{ab} & f_{bb} & f_{bc} \\
f_c & f_{ac} & f_{bc} & f_{cc}
\end{array}
\right).
\label{Hessian}
\end{equation}
The differential $dM=M_ada+M_bdb+M_cdc$ is a matrix-valued   form
\begin{gather*}
dM=
\left(\begin{array}{cccc}
0 & f_{aa} & f_{ab} & f_{ac} \\
f_{aa} & f_{aaa} & f_{aab} & f_{aac} \\
f_{ab} & f_{aab} & f_{abb} & f_{abc} \\
f_{ac} & f_{aac} & f_{abc} & f_{acc}
\end{array} \right)
da+
\left(\begin{array}{cccc}
0 & f_{ab} & f_{bb} & f_{bc} \\
f_{ab} & f_{aab} & f_{abb} & f_{abc} \\
f_{bb} & f_{abb} & f_{bbb} & f_{bbc} \\
f_{bc} & f_{abc} & f_{bbc} & f_{bcc}
\end{array} \right)
db
\\+
\left(\begin{array}{cccc}
0 & f_{ac} & f_{bc} & f_{cc} \\
f_{ac} & f_{aac} & f_{abc} & f_{acc} \\
f_{bc} & f_{abc} & f_{bbc} & f_{bcc} \\
f_{cc} & f_{acc} & f_{bcc} & f_{ccc}
\end{array}\right)
dc.
\end{gather*}
\end{thm}

Finally, we recall that the equations of gas dynamic possess only double waves reduction, and are not integrable by the method of hydrodynamic reductions \cite{FK1}. On the other hand, the generalised equations \eqref{sys_1} define a (2+1)-dimensional integrable system when the Lagrangian density $f(\varphi_x, \varphi_y, \varphi_t)$, obtained by the Hamiltonian density $h(\rho,u,v)$ performing a partial Legendre transform \eqref{legendre}, satisfies the conditions given by Theorem \ref{thm_Fer}.

\subsection{Three-component Hamiltonian systems with degenerate structure}
We have seen that the degenerate Hamiltonian operator $\eqref{rank2_P_3}_2$ leads to a class of integrable systems related to the Lagrangian density of the form $f(\varphi_x, \varphi_y, \varphi_t)$. Here we are going to describe all three-component cases arising from our classification.

The aim of this section is to apply the method of hydrodynamic reductions to three-component Hamiltonian systems given by ${\bf u}_t + P h_{\bf u}=0$, where $P$ is a Hamiltonian structure appearing in Theorems \ref{thm_2D_3cmpt_rank0}, \ref{thm_2D_3cmpt_rank1} and \ref{thm_2D_3cmpt_rank2}. Let us identify the Hamiltonian operators we obtained with the rank of the pencil $g_{\lambda}$. For instance, we call rank-zero structures the Hamiltonian operators listed in Theorem \ref{thm_2D_3cmpt_rank0}.
\begin{thm}\label{thm_sys}
The method of hydrodynamic reductions imposes additional differential constraints under which equations under study reduce to known classes of systems considered before:
\begin{itemize}
\item rank-zero structures lead to trivial systems
$$u^1_t=u^2_t=u^3_t=0,
$$
\item rank-one structures lead to one dimensional system of the form
$$u^1_t+f(u^1) u^1_x=0, \quad u^2_t=u^3_t=0,$$
\item rank-two structures lead either to one dimensional system to the form
$$
u^1_t+(h_{u^2})_x=0, \quad u^2_t+(h_{u^1})_x=0, \quad u^3_t=0,
$$
or two-component non-degenerate Hamiltonian systems 
\begin{equation}\label{2cmpt_1}
u^1_t+ (h_{u^1})_x=0,
\quad
u^2_t+ (h_{u^2})_y=0,
\end{equation}
\begin{equation}\label{2cmpt_2}
u^1_t+ (h_{u^2})_x=0,
\quad
u^2_t+(h_{u^1})_x+ (h_{u^2})_y=0,
\end{equation}
\begin{equation}\label{2cmpt_3}
u^1_t+ ( 2 u^1 h_{u^1} + u^2 h_{u^2} - h)_x + (u^1 h_{u^2})_y=0,
\quad
u^2_t+(u^2 h_{u^1})_x+( 2 u^2 h_{u^2} + u^1 h_{u^1} - h)_y=0,
\end{equation}
plus the trivial equation $u^3_t=0$, or to the system
\begin{equation}\label{3cmpt_lag}
u^1_t+ (h_{u^2})_x + (h_{u^3})_y=0,
\quad
u^2_t+ (h_{u^1})_x=0,
\quad
u^3_t+ (h_{u^1})_y=0.
\end{equation}
\end{itemize}
\end{thm}

We point out that the integrability of two-component non-degenerate Hamiltonian systems \eqref{2cmpt_1}, \eqref{2cmpt_2} and \eqref{2cmpt_3}, generated respectively by the Hamiltonian operators
$$
P=
\begin{pmatrix}
d_x & 0\\
0 & d_y
\end{pmatrix},
\
P=
\begin{pmatrix}
0 & d_x\\
d_x & d_y
\end{pmatrix},
\
P=
\begin{pmatrix}
2 u^1 &  u^2\\
u^2 & 0
\end{pmatrix}
d_x+
\begin{pmatrix}
0 & u^1\\
u^1 & 2 u^2
\end{pmatrix}
d_y+
\begin{pmatrix}
u^1_x & u^1_y\\
u^2_x & u^2_y
\end{pmatrix},
$$
is completely understood, see \cite{FOS} for further details.
Furthermore, as we showed above, system \eqref{3cmpt_lag} reduces to the three-dimensional Euler-Lagrange equations \eqref{eq_lagrange} after performing a partial Legandre transformation of the form \eqref{legendre}.

\begin{center} {\bf Proof of Theorem \ref{thm_sys}:} \end{center}\vspace{0.5pt}

First of all, let us remark that if $u^i_t=0$, for some $i$, the method of hydrodynamic reductions necessarily implies $u^i=const$.
Secondly, if one of the equations of the system is of the form
$u^i_t+\phi({\bf u}) u^i_x + \psi({\bf u}) u^i_y=0$, the method of hydrodynamic reductions implies
$(\lambda^j +\phi+ \psi \mu^j)\partial_j u^i=0 $, which leads to $u^i=const$, since we are imposing that $\lambda^j$ and $\mu^j$ do not satisfy any linear relation.
Furthermore,  in these cases we can replace $u^i$ with a constant, and then the Hamiltonian will depend on $u^j$ for $j\neq i$. 

Using these observations, the proof is straightforward. Rank-zero structures easily lead to trivial systems. For the rank-one structures we always have $u^2$ and $u^3$ constant, which leads to an operator of the form
$$
P=
\begin{pmatrix}
d_x+\kappa d_y & 0 & 0\\
0 & 0 & 0\\
0 & 0 & 0
\end{pmatrix},
\quad
\kappa=const,
$$ 
which is essentially one-dimensional (up to linear change of the independent variables $x$ and $y$).

The analysis of rank-two structures is a bit more complicated.
In the cases $\eqref{rank2_P_1}_1$ and $\eqref{rank2_P_4}_2$, the method of hydrodynamic reductions implies $u^3=const$. Thus, up to a change of local coordinates $u^1, u^2$, the $3\times3$ degenerate Hamiltonian operator reduces to direct sum of the $2\times2$ two-component non-degenerate Mokhov's Hamiltonian operator \cite{Mokhov1,Mokhov2}
$$
P=
\begin{pmatrix}
2 u^1 &  u^2\\
u^2 & 0
\end{pmatrix}
d_x+
\begin{pmatrix}
0 & u^1\\
u^1 & 2 u^2
\end{pmatrix}
d_y+
\begin{pmatrix}
u^1_x & u^1_y\\
u^2_x & u^2_y
\end{pmatrix},
$$
and the trivial $1\times 1$ operator $P=0$.

In the cases $\eqref{rank2_P_2}_{1,2}$, $\eqref{rank2_P_3}_1$, $\eqref{rank2_P_5}$ and $\eqref{rank2_P_6}$ the method of hydrodynamic reductions implies again $u^3=const$. These structures reduce to direct sum of constant $2\times2$ two-component non-degenerate Hamiltonian operator, and the trivial $1\times 1$ operator $P=0$. Constant $2\times2$ non-degenerate Hamiltonian operators are known \cite{FOS}: if they do not reduce to one-dimensional operator
$$
P=
\begin{pmatrix}
0 & d_x\\
d_x & 0
\end{pmatrix},
$$
(for instance, $\eqref{rank2_P_3}_1$ for $\epsilon=1$), they can be brought to one of the following two forms
$$
P=
\begin{pmatrix}
d_x & 0\\
0 & d_y
\end{pmatrix},
\quad
P=
\begin{pmatrix}
0 & d_x\\
d_x & d_y
\end{pmatrix},
$$
by a change of local coordinates $u^1,u^2$ and a linear change of the independent variables $x, y$.

It remains to consider the cases $\eqref{rank2_P_1}_{2}$ and $\eqref{rank2_P_4}_{1}$. It is not difficult to see that in both cases we get $u^2_y=u^3_x$. Therefore, solutions are necessarily potential. Then, setting $u^2=\varphi_x$ and $u^3=\varphi_y$, the system leads to \eqref{sys_1_new}.
\endproof

\section{Concluding remarks}
The problem of classification of 2D Hamiltonian operators of differential geometric-type, proposed by Dubrovin and Novikov in \cite{DN1}, is now completely solved up to three-component case. Even though in \cite{FLS} we provided a complete classification of non-degenerate operators up to four components, in the degenerate case it is still open.
The main obstacle is the lack of a full description of one-dimensional degenerate Poisson brackets. Indeed, already for four-component one-dimensional degenerate structures, the computation of Jacobi conditions is quite complicated \cite{Grinberg,Sav}.

As we have said above, any 2D degenerate Hamiltonian operator gives rise to a pair of 1D compatible degenerate brackets of Dubrovin-Novikov type (see \cite{Mokhov2,Mokhov3} for further details).
Some of the degenerate bi-Hamiltonian structures arising from our classification are not of the kind investigated by Strachan \cite{Strachan1,Strachan}. It would be interesting to analyse these structures and to study a possible correspondence with the analogous of Frobenius manifolds with degenerate metric.

\section*{Acknowledgments}
I would like to thank Eugene Ferapontov and Paolo Lorenzoni for useful remarks.


\section*{Appendix. Proof of Theorems \ref{thm_2D_3cmpt_rank0}, \ref{thm_2D_3cmpt_rank1} and \ref{thm_2D_3cmpt_rank2}}
\subsubsection*{Proof of Theorem \ref{thm_2D_3cmpt_rank0}}
In the case where the pencil $g_{\lambda}$ has rank constantly equal to 0, both the metrics must be identically null. Thus, by Theorem \ref{thm3cmpts}, we can always reduce the coefficients $b^{ij}_k$ to $b^{12}_3=-b^{21}_3=1$, that is, the $x$-part of the 2D Hamiltonian operator can be fixed as \eqref{rank0}.
Imposing \eqref{mok_all}, we get that all the coefficients $\tilde b^{ij}_k$ must vanish except $\tilde b^{12}_3=-\tilde b^{21}_3=\nu(u^1,u^2,u^3)$.
The transformations which preserve the form of $P_{(x)}$ have the form
$$
u^1=\varphi^1(v^1,v^2,v^3), \quad u^2=\varphi^2(v^1,v^2,v^3), \quad u^3=\varphi^3(v^3),
$$
with the constraint
$$
\partial_1 \varphi^1 \partial_2  \varphi^2 - \partial_2  \varphi^1 \partial_1  \varphi^2=(\varphi^3)'.
$$

If $\nu=\xi$ is constant, we get $\tilde b=\xi b$. Let us assume that $\nu=\nu(u^3)$. Then, in the new system of coordinates, it is always possible to reduce $\nu$ to $v^3$. Let us finally assume that $\nu$ is an arbitrary function of $u^1,u^2,u^3$. Then, there exists a change of coordinate preserving $P_{(x)}$ which transforms $\nu$ to $v^1$ or, equivalently, to $v^2$ (these two cases are the same since we can swap $v^1$, $v^2$). Summarising, $P_{(y)}$ leads to one of the following two structures
$$
P_{(y)}=
\begin{pmatrix}
0 & v^1 v^3_y & 0\\
-v^1 v^3_y & 0 & 0\\
0 & 0 & 0
\end{pmatrix},
\quad
P_{(y)}=
\begin{pmatrix}
0 & v^3 v^3_y & 0\\
-v^3 v^3_y & 0 & 0\\
0 & 0 & 0
\end{pmatrix},
$$
which are not equivalent modulo transformations which preserve the form of $P_{(x)}$.


\subsubsection*{Proof of Therem \ref{thm_2D_3cmpt_rank1}}

When the rank of the pencil $g^{ij}+\lambda \tilde g^{ij}$ is constantly equal to one, we can always work in a coordinate system where the metrics assume the forms
$$
g^{ij}=
\begin{pmatrix}
1 & 0 & 0\\
0 & 0 & 0\\
0 & 0 & 0
\end{pmatrix},
\quad
\tilde g^{ij}=
\begin{pmatrix}
f & 0 & 0\\
0 & 0 & 0\\
0 & 0 & 0
\end{pmatrix},
$$
where $f=f(u^1,u^2,u^3)$. Let us now consider separately each case given by Theorem \ref{thm3cmpts}.

\medskip \noindent
{\bf Case} $\eqref{rank1}_1$. The symbols $b^{ij}_k$ are identically $0$. In this case, a generic transformation which preserves $P_{(x)}$ is given by
\begin{equation}\label{change3cmp_rk1}
u^1=v^1+\varphi^1(v^2,v^3), \quad u^2=\varphi^2(v^2,v^3), \quad u^3=\varphi^3(v^2, v^3).
\end{equation}
Let us point out that this change of coordinates transforms $b^{ij}_k$ (and then $\tilde b^{ij}_k$) as components of a $(2,1)$-tensor \cite{Sav}.
Conditions \eqref{mok_all} imply two solutions.

\medskip \noindent
{\it Solution 1}. The first solution reads
$$
f=f(u^2,u^3), \quad
\tilde b^{11}_2=\frac{\partial_2 f}{2}, \quad \tilde b^{11}_3=\frac{\partial_3 f}{2},
\quad
\tilde b^{21}_2=-\tilde b^{12}_2 =\tilde b^{13}_3=-\tilde b^{31}_3=\psi,
$$ $$
\tilde b^{21}_3=-\tilde b^{12}_3=\frac{\psi^2}{\eta},
\quad
\tilde b^{13}_2=-\tilde b^{31}_2=\eta,
$$
where $\psi=\psi(u^2,u^3)$ and $\eta=\eta(u^2,u^3)$, and all other $b^{ij}_k$ vanish. Clearly, we are imposing $\eta\neq0$. The operator leads to
$$
P_{(y)}=
\begin{pmatrix}
f d_y +\frac{\partial_2 f u^2_y + \partial_3 f u^3_y}{2}& -\psi u^2_y -\frac{\psi^2}{\eta} u^3_y & \eta u^2_y + \psi u^3_y\\
\psi u^2_y +\frac{\psi^2}{\eta} u^3_y & 0 & 0\\
-\eta u^2_y - \psi u^3_y & 0 & 0
\end{pmatrix},
$$
where $f=f(u^2,u^3)$, $\psi=\psi(u^2,u^3)$ and $\eta=\eta(u^2,u^3)\neq0$. Applying a transformation of the form \eqref{change3cmp_rk1} we get
$f  \rightarrow   f(\varphi^2,\varphi^3)$ and
\begin{gather*}
\tilde b^{21}_2=-\tilde b^{12}_2 =\tilde b^{13}_3=-\tilde b^{31}_3=\psi
\quad \rightarrow \quad
\frac{(\partial_3 \varphi^2 \eta + \partial_3 \varphi^3 \psi)(\partial_2 \varphi^2 \eta + \partial_2 \varphi^3 \psi)}{(\partial_2 \varphi^2 \partial_3 \varphi^3-\partial_3 \varphi^2 \partial_2 \varphi^3)\eta}, \\
\tilde b^{21}_3=-\tilde b^{12}_3=\frac{\psi^2}{\eta}
\quad \rightarrow \quad
-\frac{(\partial_3 \varphi^2 \eta + \partial_3 \varphi^3 \psi)^2}{(\partial_2 \varphi^2 \partial_3 \varphi^3-\partial_3 \varphi^2 \partial_2 \varphi^3)\eta},\\
\tilde b^{13}_2=-\tilde b^{31}_2=\eta
\quad \rightarrow \quad
\frac{(\partial_2 \varphi^2 \eta + \partial_2 \varphi^3 \psi)^2}{(\partial_2 \varphi^2 \partial_3 \varphi^3-\partial_3 \varphi^2 \partial_2 \varphi^3)\eta}.
\end{gather*}
We cannot choose both $\partial_2 \varphi^2 \eta + \partial_2 \varphi^3 \psi=0$ and $\partial_3 \varphi^2 \eta + \partial_3 \varphi^3 \psi=0$, otherwise we would have the denominator equal to zero. However, a suitable choice of the functions $\varphi^2$ and $\varphi^3$ allows us to reduce $f$ to either $v^2$ or $v^3$ (which are equivalent up to swapping $v^2$ and $v^3$) if $f$ is not constant, and $\psi$ to zero. This leads to two operators
$$
P_{(y)}=
\begin{pmatrix}
\kappa d_y & 0 & \tilde{\eta} v^2_y \\
0 & 0 & 0\\
-\tilde{\eta} v^2_y & 0 & 0
\end{pmatrix},
\quad
P_{(y)}=
\begin{pmatrix}
v^2 d_y +\frac{v^2_y}{2}& 0 & \tilde{\eta} v^2_y \\
0 & 0 & 0\\
-\tilde{\eta} v^2_y & 0 & 0
\end{pmatrix},
$$
where $\kappa$ is constant and $\tilde{\eta}=\tilde{\eta}(v^2,v^3)$. Allowing linear change of $x$ and $y$, $\kappa$ can be brought to zero.

\medskip \noindent
{\it Solution 2}. In the case where $\eta=0$, the solution reads
$$
f=f(u^2,u^3), \quad
\tilde b^{11}_2=\frac{\partial_2 f}{2}, \quad \tilde b^{11}_3=\frac{\partial_3 f}{2},
\quad
\tilde b^{12}_3=-\tilde b^{21}_3 =\nu(u^2,u^3),
$$
and all other $b^{ij}_k$ vanish.
Modulo transformations of the form \eqref{change3cmp_rk1} this case corresponds to the previous one.

\medskip \noindent
{\bf Case} $\eqref{rank1}_2$. Here $b^{12}_3=-b^{21}_3=1$, while other symbols $b^{ij}_k$ are identically $0$.
Conditions \eqref{mok_all} imply
$$
f=f(u^2,u^3), \quad
\tilde b^{11}_2=\frac{\partial_2 f}{2}, \quad \tilde b^{11}_3=\frac{\partial_3 f}{2},
\quad
\tilde b^{12}_3=-\tilde b^{21}_3=\nu(u^2,u^3),
$$
and other $\tilde b^{ij}_k=0$.
A generic transformation which preserves $P_{(x)}$ is given by
$$
u^1=v^1, \quad u^2=\partial_3 \varphi^3(v^3)v^2+ \varphi^2(v^3), \quad u^3=\varphi^3(v^3).
$$
Unfortunately, in general this group of transformations cannot help to simplify our structure (the operator depends on two functions of $u^2, u^3$, while the group depends only on two functions of $u^3$). We should consider separately each case where the functions $f$ and $\nu$ are constant or depend on one single variables. Therefore, it is more reasonable to consider just the general solution, namely
$$
P_{(y)}=
\begin{pmatrix}
f d_y +\frac{\partial_2 f u^2_y + \partial_3 f u^3_y}{2}& \nu u^3_y & 0\\
-\nu  u^3_y & 0 & 0\\
0 & 0 & 0
\end{pmatrix},
$$
for arbitrary $f=f(u^2,u^3)$, $\nu=\nu(u^2,u^3)$.


\medskip \noindent
{\bf Case} $\eqref{rank1}_3$. Here $b^{31}_3=-b^{13}_3=\frac{1}{u^1}$, while other symbols $b^{ij}_k$ are identically $0$.
Conditions \eqref{mok_all} imply
$$
f=f(u^2,u^3), \quad
\tilde b^{11}_2=\frac{\partial_2 f}{2}, \quad \tilde b^{11}_3=\frac{\partial_3 f}{2},
\quad
\tilde b^{31}_3=-\tilde b^{13}_3=\frac{f}{u^1},
\quad
\tilde b^{13}_2=-\tilde b^{31}_2=\frac{\nu}{u^1},
$$
where $\nu=\nu(u^2,u^3)$, and other $\tilde b^{ij}_k=0$.
A generic transformation which preserves $P_{(x)}$ is given by
$$
u^1=v^1, \quad u^2=\varphi^2(v^2), \quad u^3=\varphi^3(v^3).
$$
As before, this group of transformations cannot help to simplify our structure for arbitrary $f$ and $\nu$. Therefore, the operator leads to
$$
P_{(y)}=
\begin{pmatrix}
f d_y +\frac{\partial_2 f u^2_y +\partial_3 f u^3_y }{2}& 0 & \frac{\nu u^2_y -f u^3_y}{u^1}\\
0 & 0 & 0\\
-\frac{\nu u^2_y -f u^3_y}{u^1} & 0 & 0
\end{pmatrix},
$$
for arbitrary $f=f(u^2,u^3)$, $\nu=\nu(u^2,u^3)$.


\medskip \noindent
{\bf Case} $\eqref{rank1}_4$. Here $b^{21}_2=-b^{12}_2=-b^{13}_3=b^{31}_3=\frac{1}{u^1}$, while other symbols $b^{ij}_k$ are identically $0$. Conditions \eqref{mok_all} imply
$$
f=f(u^2,u^3), \quad
\tilde b^{11}_2=\frac{\partial_2 f}{2}, \quad \tilde b^{11}_3=\frac{\partial_3 f}{2},
\quad
\tilde b^{21}_2=-\tilde b^{12}_2=-\tilde b^{13}_3=\tilde b^{31}_3=\frac{f}{u^1},
$$
and other $\tilde b^{ij}_k=0$. A generic transformation which preserves $P_{(x)}$ is given by
$$
u^1=v^1, \quad u^2=\varphi^2(v^2,v^3), \quad u^3=\varphi^3(v^2,v^3).
$$
This change of coordinates transforms the objects $\tilde b^{ij}_k$ as components of a $(2,1)$-tensor \cite{Sav}. If $f=\xi$ is constant, then $\tilde g=\xi g$ and $\tilde b=\xi b$. Otherwise, we can choose $\varphi^2$ or $\varphi^3$ such that $f$ reduce either to $v^2$ or $v^3$, which are equivalent forms since we can swap $v^2$, $v^3$. Thus $P_{(y)}$ leads to 
$$
P_{(y)}=
\begin{pmatrix}
v^2 d_y +\frac{v^2_y}{2} & -\frac{v^2 v^2_y}{v^1} & -\frac{v^2 v^3_y}{v^1} \\
\frac{v^2 v^2_y}{v^1}  & 0 & 0\\
\frac{v^2 v^3_y}{v^1}  & 0 & 0
\end{pmatrix}.
$$


\subsubsection*{Proof of Therem \ref{thm_2D_3cmpt_rank2}}
When the rank of the pencil $g^{ij}+\lambda \tilde g^{ij}$ is constantly equal to two, we have three possibilities:
\begin{equation}\label{rk2form1}
g^{ij}=
\begin{pmatrix}
0 &1&0\\
1&0&0\\
0&0&0
\end{pmatrix},
\quad
\tilde g^{ij}=
\begin{pmatrix}
p &q&0\\
q&r&0\\
0&0&0
\end{pmatrix},
\end{equation}
\begin{equation}\label{rk2form2}
g^{ij}=
\begin{pmatrix}
0 &1&0\\
1&0&0\\
0&0&0
\end{pmatrix},
\quad
\tilde g^{ij}=
\begin{pmatrix}
p &q&r\\
q&0&0\\
r&0&0
\end{pmatrix},
\end{equation}
or
\begin{equation}\label{rk2form3}
g^{ij}=
\begin{pmatrix}
0 &1&0\\
1&0&0\\
0&0&0
\end{pmatrix},
\quad
\tilde g^{ij}=
\begin{pmatrix}
0 &q&0\\
q& p&r\\
0&r&0
\end{pmatrix},
\end{equation}
where $p,q,r$ are arbitrary functions of $u^1, u^2, u^3$. We remark that \eqref{rk2form2} and \eqref{rk2form3} are equivalent up to a transformation of the form $u^1=v^2, u^2=v^1, u^3=v^3$ (which preserves the form of the first metric). However, this change of coordinate does not fix the structures $\eqref{rank2}_2$ and $\eqref{rank2}_3$. Therefore, when the $x$-part of the operator is given by $\eqref{rank2}_1$, we can avoid \eqref{rk2form3}, while in the other two cases, $\eqref{rank2}_2$ and $\eqref{rank2}_3$, we have to take it into account.

\medskip \noindent
{\bf Case} $\eqref{rank2}_1$. Here the first structure is $\eqref{rank2}_1$, and the group of transformations which preserve its form is given by
\begin{equation}\label{group1}
u^1=e^{\psi} v^1+\varphi^1(v^3), \quad u^2=e^{-\psi} v^2+\varphi^2(v^3), \quad u^3=\varphi^3(v^3),
\end{equation}
where $\psi$ is constant, plus the switch of $u^1$, $u^2$ (note that this change of coordinates transforms the objects $\tilde b^{ij}_k$ as components of a $(2,1)$-tensor \cite{Sav}). In the case where we are dealing with \eqref{rk2form1}, up to swapping $u^1, u^2$, solutions of conditions \eqref{mok_all} can be summarised as follows.

\medskip \noindent
{\it Solution 1}. The first solution is given by
\begin{gather*}
p=p(u^3)+\kappa u^1, \quad q=q(u^3) -\frac{\kappa u^2}{2}, \quad
\tilde b^{11}_1=\tilde b^{21}_2=\frac{\kappa}{2}, \quad \tilde b^{11}_3=\frac{p'}{2},
\\
\tilde b^{12}_2=  -\kappa, \quad \tilde b^{12}_3=2 q', \quad \tilde b^{21}_3=-q',
\end{gather*}
where $\kappa\neq0$ is constant. This leads to
$$
P_{(y)}=
\begin{pmatrix}
(p+\kappa u^1) \ d_y +\frac{\kappa u^1_y +p' u^3_y}{2}& \left(q -\frac{\kappa u^2}{2}\right) \ d_y -\kappa u^2_y + 2 q' u^3_y &0\\
\left(q -\frac{\kappa u^2}{2}\right) \ d_y +\frac{\kappa u^2_y}{2}-q' u^3_y& 0 & 0\\
0 & 0 & 0
\end{pmatrix}.
$$
A change of coordinates of the form \eqref{group1} transforms
$$
p \ \to \ \varphi^1 \kappa+p(\varphi^3), \quad q \ \to \ -\frac{\varphi^2 \kappa}{2}+q(\varphi^3), \quad \kappa \ \to \kappa e^{-\psi}
$$
thus, it is always possible to choose $\varphi^1$, $\varphi^2$ and $\psi$ such that in the new coordinates $p=q=0$ and $\kappa$ is fixed, let us set it equal to $-2$. Therefore, the operator leads to
$$
P_{(y)}=
\begin{pmatrix}
-2 v^1 \ d_y -v^1_y&  v^2 \ d_y +2 v^2_y &0\\
v^2 \ d_y - v^2_y& 0 & 0\\
0 & 0 & 0
\end{pmatrix}.
$$
Let us point out that in this case the 2D operator $P$ can be view as direct sum of $2\times2$ Mokhov's operator \cite{Mokhov2}
$$
P=
\begin{pmatrix}
-2 v^1 \ d_y -v^1_y&  d_x+v^2 \ d_y +2 v^2_y \\
d_x+ v^2 \ d_y - v^2_y& 0 
\end{pmatrix}.
$$
and trivial $1\times 1$ operator $P=0$.

\medskip \noindent
{\it Solution 2}. The second solution is given by
\begin{gather*}
p=p(u^3)+\kappa u^1, \quad q=q(u^3) -\frac{\kappa u^2}{2}, \quad
\tilde b^{11}_1=\tilde b^{21}_2=\tilde b^{31}_3=-\tilde b^{13}_3=\frac{\kappa}{2}, \quad \tilde b^{11}_3=\frac{p'}{2},
\\
\tilde b^{12}_2=  -\kappa, \quad \tilde b^{12}_3= q',
\end{gather*}
where $\kappa\neq0$ is constant. This leads to
$$
P_{(y)}=
\begin{pmatrix}
(p+\kappa u^1) \ d_y +\frac{\kappa u^1_y +p' u^3_y}{2}& \left(q -\frac{\kappa u^2}{2}\right) \ d_y -\kappa u^2_y + q' u^3_y & -\frac{\kappa u^3_y}{2}\\
\left(q -\frac{\kappa u^2}{2}\right) \ d_y +\frac{\kappa u^2_y}{2}& 0 & 0\\
\frac{\kappa u^3_y}{2} & 0 & 0
\end{pmatrix}.
$$
The group \eqref{group1} acts on this case as the previous one. Thus, we can reduce $p$ and $q$ to zero, and $\kappa$ to $-2$, obtaining
$$
P_{(y)}=
\begin{pmatrix}
-2 v^1\ d_y -v^1_y & v^2 \ d_y +2 v^2_y  &  v^3_y\\
 v^2 \ d_y -v^2_y& 0 & 0\\
-v^3_y& 0 & 0
\end{pmatrix}.
$$
\medskip \noindent
{\it Solution 3}. In the case where $\kappa=0$, the solution is given by
\begin{gather*}
p=p(u^3), \quad q=q(u^3) , \quad r=r(u^3), \quad
\tilde b^{11}_3=\frac{p'}{2}, \quad \tilde b^{12}_3= \nu(u^3), \quad \tilde b^{21}_3=q'-\nu,
\quad \tilde b^{22}_3=\frac{r'}{2}.
\end{gather*}
This leads to
$$
P_{(y)}=
\begin{pmatrix}
p\ d_y +\frac{p' u^3_y}{2}&q \ d_y +\nu u^3_y & 0\\
q\ d_y +(q'-\nu) u^3_y& r \ d_y +\frac{r' u^3_y}{2} & 0\\
0 & 0 & 0
\end{pmatrix}.
$$
Here, the group \eqref{group1} acts on the objects as
$$
p \ \to \ p(\varphi^3) e^{-2\psi}, \quad q \ \to \ q(\varphi^3), \quad r \ \to \ r(\varphi^3) e^{2\psi},
\quad \nu \to (\varphi^3)'  \nu(\varphi^3).
$$
Here we have four arbitrary functions $p,q,r,\nu$ and only one function $\varphi^3$ and one constant $\psi$ acting on them. Thus, even if we could consider several cases (where some functions are constant or zero), it does not simplify the classification. However, let us make a choice: if $\nu$ is non-zero, it can be always reduced to 1. Thus the operator leads to
\begin{equation}\label{op_3_rank2}
P_{(y)}=
\begin{pmatrix}
p\ d_y +\frac{p' u^3_y}{2}&q \ d_y +\epsilon u^3_y & 0\\
q\ d_y +(q'-\epsilon) u^3_y& r \ d_y +\frac{r' u^3_y}{2} & 0\\
0 & 0 & 0
\end{pmatrix},
\end{equation}
with $\epsilon$ equal either to $0$ or $1$.

\medskip
Let us deal with \eqref{rk2form2}. If $r=0$, solutions of conditions \eqref{mok_all} lead to \eqref{op_3_rank2} (replacing $r$ with 0). Otherwise, we have
$$
p=p(u^3), \quad q=(u^3), \quad r=r(u^3), \quad \tilde b^{11}_3=\frac{p'}{2}, \quad \tilde b^{12}_3=q', \quad \tilde b^{13}_3= r',
$$
which leads to
$$
P_{(y)}=
\begin{pmatrix}
p \ d_y + \frac{p' u^3_y}{2} & q \ d_y +q' u^3_y & r \ d_y + r' u^3_y\\
q \ d_y &0 & 0\\
r \ d_y  & 0 & 0
\end{pmatrix}.
$$
In this case, \eqref{group1} transform $p,q,r$ as
$$
p \ \to \ e^{-2\psi} \left(p(\varphi^3)-\frac{2 (\varphi^1)' r(\varphi^3)}{(\varphi^3)'}\right) ,
\quad q \ \to \ q(\varphi^3)-\frac{(\varphi^2)' r(\varphi^3)}{(\varphi^3)'},
\quad r \ \to \ \frac{r(\varphi^3)}{(\varphi^3)'} e^{2\psi}.
$$
Thus, since $r\neq0$, we can always reduce $r$ to $1$ and $p$ and $q$ to $0$, obtaining
$$
P_{(y)}=
\begin{pmatrix}
0 & 0& d_y \\
0 &0 & 0\\
d_y  & 0 & 0
\end{pmatrix}.
$$

\noindent
{\bf Case} $\eqref{rank2}_2$. Here the first structure is $\eqref{rank2}_2$, and the group of transformations which preserve its form is given by
\begin{equation}\label{change_rk2_case2}
u^1=e^{\psi(v^3)} v^1+\varphi^1(v^3), \quad u^2=e^{-\psi(v^3)} v^2, \quad u^3=\varphi^3(v^3).
\end{equation}

In the case where the second metric is of the form \eqref{rk2form1}, conditions \eqref{mok_all} lead to two structures, given respectively by
\begin{equation}\label{rank2_2D_1}
p=p(u^3), \quad q=q(u^3), \quad r=0, \quad \tilde b^{11}_3=\frac{p'}{2}, \quad \tilde b^{13}_3=-\tilde b^{31}_3=\frac{q}{u^2}, \quad \tilde b^{21}_3=q',
\end{equation}
and
\begin{equation}\label{rank2_2D_2}
\begin{array}{c}
p=p(u^3) u^1+\tilde p(u^3), \quad q=\kappa-\dfrac{p u^2}{2},  \quad r=0, \quad \tilde b^{11}_1=\tilde b^{21}_2=\dfrac{p}{2},
\quad \tilde b^{11}_3=\dfrac{p' u^1 +\tilde p'}{2}, \quad \tilde b^{12}_2=-p,
\\[5pt]
\tilde b^{12}_3=-\dfrac{p' u^2}{2},
\quad \tilde b^{31}_3=-\tilde b^{13}_3=\dfrac{\kappa}{u^2},
\end{array}
\end{equation}
where $\kappa$ is constant.

In the case where the second structure is of the form given by \eqref{rk2form2}, conditions \eqref{mok_all} lead again to two structures, where the first is the same as \eqref{rank2_2D_1}, and the second is given by
\begin{gather*}
p=p(u^3) u^1+\tilde p(u^3), \quad q=\kappa-\frac{p u^2}{2},  \quad r=r(u^3), \quad \tilde b^{11}_1=\tilde b^{21}_2=\frac{p}{2},
\quad \tilde b^{11}_3=\frac{p' u^1 +\tilde p'}{2}, \quad \tilde b^{12}_2=-p,
\\
\tilde b^{12}_3=-\frac{p' u^2}{2},
\quad \tilde b^{13}_3=-\tilde b^{31}_3=r-\frac{\kappa}{u^2},
\quad
\quad \tilde b^{13}_2=-\tilde b^{31}_2=\frac{r}{u^2},
\end{gather*}
which corresponds to \eqref{rank2_2D_2} replacing $r, \tilde b^{13}_2, \tilde b^{13}_3, \tilde b^{31}_2, \tilde b^{31}_3$ with
$$
r=r(u^3), \quad \tilde b^{13}_2=-\tilde b^{31}_2=\frac{r}{u^2}, \quad 
\tilde b^{13}_3=-\tilde b^{31}_3=r-\frac{\kappa}{u^2}.
$$

Finally, considering \eqref{rk2form3}, conditions \eqref{mok_all} imply
$$
p=0, \quad q=q(u^3), \quad r=r(u^3), \quad \tilde b^{13}_1=-\tilde b^{31}_1=\frac{r}{u^2}, \quad \tilde b^{31}_3=-\tilde b^{13}_3=\frac{q}{u^2}, \quad \tilde b^{21}_3=q', \quad \tilde b^{23}_3=r'.
$$

Summarising, this case leads to three different structures, namely
\begin{equation}\label{rk2_2D_1}
P_{(y)}=
\begin{pmatrix}
p \ d_y +\frac{p' u^3_y}{2}& q \ d_y & -\frac{q u^3_y}{u^2} \\
q \ d_y +q' u^3_y& 0 & 0\\
\frac{q u^3_y}{u^2}&0 &0
\end{pmatrix},
\quad
P_{(y)}=
\begin{pmatrix}
0 & q \ d_y & \frac{r u^1_y - q u^3_y}{u^2} \\
q \ d_y + q' u^3_y & 0 & r \ d_y + r' u^3_y\\
\frac{q u^3_y-r u^1_y}{u^2} & r \ d_y & 0
\end{pmatrix}.
\end{equation}
\begin{equation}\label{rk2_2D_2}
P_{(y)}=
\begin{pmatrix}
(p u^1 +\tilde p)\ d_y +\frac{p u^1_y +(p' u^1 + \tilde p')u^3_y}{2}& \left(\kappa-\frac{p u^2}{2} \right) \ d_y -p u^2_y-\frac{p' u^2 u^3_y}{2}& r \ d_y + \frac{r u^2_y + (r' u^2 -\kappa)u^3_y}{u^2} \\
\left(\kappa-\frac{p u^2}{2} \right)  \ d_y +\frac{p u^2_y}{2}& 0 & 0\\
r \ d_y+\frac{\kappa u^3_y-r u^2_y}{u^2}&0 &0
\end{pmatrix}.
\end{equation}
These are the more general solutions assuming the first operator given by $\eqref{rank2}_2$. In these cases, a transformation of the form \eqref{change_rk2_case2} allows to simplify these structures, but we will get more cases. Let us discuss each operator in detail.

Let us consider an operator of the form $\eqref{rk2_2D_1}_1$. Under change of coordinates of the form \eqref{change_rk2_case2}, $p$ and $q$ transform as
$$
p \ \to \ e^{-2 \psi} p(\varphi^3), \quad q \ \to \ q(\varphi^3).
$$
Thus, if $p$ vanishes and $q=\xi$ is constant, we get $\tilde g= \xi g$, $\tilde b =\xi b$, that is, a trivial operator. If $p$ vanishes, but $q$ is arbitrary, it can be easily reduce to $v^3$. Otherwise, if $p\neq 0$, it can be always reduced to $1$, and, if $q$ is not constant, the freedom in $u^3=\varphi^3(v^3)$ allows to reduce $q$ to $v^3$. Therefore, in this case we get the following non-trivial canonical forms
$$
P_{(y)}=
\begin{pmatrix}
\epsilon d_y & v^3 \ d_y & -\frac{v^3 v^3_y}{v^2} \\
v^3 \ d_y + v^3_y& 0 & 0\\
\frac{v^3 v^3_y}{v^2}&0 &0
\end{pmatrix},
\quad
P_{(y)}=
\begin{pmatrix}
d_y & \kappa \ d_y & -\frac{\kappa v^3_y}{v^2} \\
\kappa \ d_y & 0 & 0\\
\frac{\kappa v^3_y}{v^2}&0 &0
\end{pmatrix},
$$
where $\kappa$ is constant and $\epsilon$ can be either 0 or 1.

\medskip
When the operator takes the form $\eqref{rk2_2D_1}_2$, if $r$ vanishes, it reduces to the first operator of the previous case with $\epsilon=0$. In general, a change of coordinates given by \eqref{change_rk2_case2} with $\psi=const$, transforms $r$ and $q$ as
$$
r \ \to \ \frac{r(\varphi^3)}{(\varphi^3)'}, \quad q \ \to \ q(\varphi^3)- \frac{r(\varphi^3)(\varphi^1)'}{(\varphi^3)'}.
$$
Notice that here we have to impose the constraint $\psi=const$, otherwise we would have an extra function in the metric written in the new coordinates.
Thus, for $r\neq0$, choosing $\varphi^1$, $\varphi^3$ such that $(\varphi^3)'=r(\varphi^3)$, $(\varphi^1)'=q(\varphi^3)$, $r$ can be brought to 1 and $q$ to 0, obtaining
$$
P_{(y)}=
\begin{pmatrix}
0 & 0 & \frac{v^1_y}{v^2} \\
0 & 0 & d_y \\
-\frac{v^1_y}{v^2} & d_y & 0
\end{pmatrix}.
$$

Finally, we have to look at the case \eqref{rk2_2D_2}. In the general case, a change of coordinates of the form \eqref{change_rk2_case2} transforms the functions $p$, $\tilde p$ and $r$ as
$$
p\  \rightarrow \ \left(p(\varphi^3)  - \frac{2 r(\varphi^3) \psi'}{(\varphi^3)'} \right) e^{-\psi},
\;
\tilde p \ \rightarrow \ \left( \tilde{p}(\varphi^3) +p(\varphi^3)\varphi^1 - \frac{2 r(\varphi^3) (\varphi^1)'}{(\varphi^3)'}\right) e^{-2\psi},
\;
r\ \rightarrow \ \frac{r(\varphi^3) e^{-\psi}}{(\varphi^3)'}.
$$
Thus, if $r\neq0$, we can reduce $p$ and $\tilde p$ to zero and $r$ to $1$. Otherwise, if $r=0$ and $p\neq0$, we can brought $p$ to 1 and $\tilde p$ to zero. If both $r$ and $p$ are equal to zero, then $\tilde p$ can be reduced to 1. Finally, if $r=p=\tilde p=0$, we have $\tilde g= \kappa g$, $\tilde b =\kappa b$.
Therefore, the canonical form of \eqref{rk2_2D_2} can be summarised as follow
$$
P_{(y)}=
\begin{pmatrix}
0 & \kappa \ d_y & d_y + \frac{v^2_y  -\kappa v^3_y}{v^2} \\
\kappa\ d_y & 0 & 0\\
d_y+\frac{\kappa v^3_y- v^2_y}{v^2}&0 &0
\end{pmatrix},
\quad
P_{(y)}=
\begin{pmatrix}
d_y &\kappa \ d_y &-\frac{\kappa v^3_y}{v^2} \\
\kappa  \ d_y & 0 & 0\\
\frac{\kappa v^3_y}{v^2}&0 &0
\end{pmatrix},
$$
$$
P_{(y)}=
\begin{pmatrix}
v^1\ d_y +\frac{v^1_y}{2}& \left(\kappa-\frac{v^2}{2} \right) \ d_y - v^2_y & -\frac{\kappa v^3_y}{v^2}\\
\left(\kappa-\frac{ v^2}{2} \right)  \ d_y +\frac{v^2_y}{2}& 0 & 0\\
\frac{\kappa v^3_y}{v^2}&0 &0
\end{pmatrix}.
$$

\begin{rmk}
Allowing linear change of the independent variables  $x$ and $y$, one can easily see that it is always possible to reduce to zero the part of $P_{(y)}$ which is proportional to $P_{(x)}$. This means that we can set $\kappa$ equal to zero in the above canonical forms.
\end{rmk}


\noindent
{\bf Case} $\eqref{rank2}_3$. Here the first structure is given by $\eqref{rank2}_3$. In the first case, that is, when the second metric is given by \eqref{rk2form1}, conditions \eqref{mok_all} imply
$$
p=p(u^3), \quad q=\kappa-u^3 p, \quad r=(u^3)^2 p,
\quad
\tilde b^{11}_3=\frac{p'}{2}, \quad
\tilde b^{12}_3=-\frac{u^3 p'}{2}, \quad
\tilde b^{31}_3=-\tilde b^{13}_3=\frac{2 u^3 p-\kappa}{u^1 u^3 - u^2},
$$
$$
\tilde b^{21}_3=-p-\frac{u^3 p'}{2},\quad
\tilde b^{22}_3=u^3 p+\frac{(u^3)^2 p'}{2}, \quad
\tilde b^{23}_3=-\tilde b^{32}_3=\frac{u^3(2 u^3p-\kappa)}{u^1 u^3 - u^2}.
$$
Thus, the operator leads
$$
P_{(y)}=
\begin{pmatrix}
p\, d_y+\frac{p' u^3_y}{2} &(\kappa - u^3 p ) \,  d_y-\frac{u^3 p' u^3_y}{2} & -\frac{(2 u^3 p -\kappa)u^3_y }{u^3 u^1-u^2}\\
(\kappa - u^3 p )\,  d_y-\frac{(2p +u^3 p') u^3_y}{2}&  (u^3)^2 p \, d_y+\frac{(2u^3p +(u^3)^2p')u^3_y }{2}& \frac{u^3(2 u^3 p -\kappa) u^3_y }{u^3 u^1-u^2}\\
\frac{(2 u^3 p -\kappa)u^3_y }{u^3 u^1-u^2}&-\frac{u^3(2 u^3 p -\kappa)u^3_y }{u^3 u^1-u^2}&0
\end{pmatrix},
$$
where $p=p(u^3)$ and $\kappa$ is constant. First of all, by linear change of $x$ and $y$, we can set $\kappa$ to $0$. A change of coordinates which preserves $P_{(x)}$, namely
$$
u^1=\sqrt{\frac{v^3}{\varphi(v^3)}}v^1+\frac{c}{\sqrt{\varphi(v^3)}}, \quad u^2=\sqrt{\frac{\varphi(v^3)}{v^3}}v^2+c \sqrt{\varphi(v^3)}, \quad u^3=\varphi(v^3),
$$
where $c=const$, transforms $p$ into $\frac{\varphi p(\varphi)}{v^3}$. This means that, if $p\neq \frac{\xi}{u^3}$ for $\xi$ constant, we can always reduce $p$ to $1$, obtaining
$$
P_{(y)}=
\begin{pmatrix}
d_y &- v^3  \,  d_y& -\frac{2 v^3 v^3_y }{v^3 v^1-v^2}\\
 - v^3 \,  d_y-v^3_y&  (v^3)^2  \, d_y+v^3 v^3_y & \frac{2 (v^3)^2  v^3_y }{v^3 v^1-v^2}\\
\frac{2 v^3 v^3_y }{v^3 v^1-v^2}&-\frac{2 (v^3)^2 v^3_y }{v^3 v^1-v^2}&0
\end{pmatrix},
$$
Otherwise, setting $p= \frac{\xi}{u^3}$, we get
$$
P_{(y)}=
\begin{pmatrix}
\frac{\xi \, d_y}{u^3}-\frac{\xi u^3_y}{(u^3)^2}& - \xi d_y +\frac{\xi u^3_y}{2 u^3}& -\frac{2 \xi u^3_y }{u^3 u^1-u^2}\\
- \xi \,d_y-\frac{\xi u^3_y}{2 u^3}&  \xi u^3 \, d_y +\frac{u^3_y}{2} & \frac{2 \xi u^3 u^3_y }{u^3 u^1-u^2}\\
\frac{2 \xi u^3_y }{u^3 u^1-u^2}&-\frac{2 \xi u^3 u^3_y }{u^3 u^1-u^2}&0
\end{pmatrix}.
$$

\medskip
In the cases where the second metric is given by \eqref{rk2form2} or by \eqref{rk2form3}, conditions \eqref{mok_all} imply
$$
p=r=0, \quad q=\kappa, \quad \tilde b^{13}_3=-\tilde b^{31}_3=\frac{\kappa}{u^1u^3-u^2},
\quad \tilde b^{23}_3=-\tilde b^{32}_3=-\frac{u^3\kappa}{u^1u^3-u^2},
$$
where $\kappa$ is constant. This leads to a trivial operator, since $\tilde g= \kappa g$ and $\tilde b=\kappa b$.



\addcontentsline{toc}{section}{References}
\begin{thebibliography}{99}


\bibitem{B1} O.I. Bogoyavlenskij, {\it Invariant foliations for the Poisson brackets of hydrodynamic type}. Phys. Lett. A {\bf 360} (2007), no. 4-5, 539-544.

\bibitem{B2} O.I. Bogoyavlenskij, {\it Tensor invariants of the Poisson brackets of hydrodynamic type}. Comm. Math. Phys. {\bf 277} (2008), no. 2, 369-384.

\bibitem{BF} C.P. Boyer and J.D. Finley, {\it Killing vectors in self-dual Euclidean Einstein spaces}. J. Math. Phys. {\bf 23} (1982), 1126-1130

\bibitem{DN}
B.A. Dubrovin and S.P. Novikov, {\it Hamiltonian formalism of one-dimensional systems of the hydrodynamic type and the Bogolyubov-Whitham averaging method}. (Russian) Dokl. Akad. Nauk SSSR {\bf 270} (1983), no. 4, 781-785. 

\bibitem{DN1} B.A. Dubrovin and S.P.  Novikov,  \emph{Poisson brackets of hydrodynamic type}. Dokl. Akad. Nauk SSSR {\bf 279}, (1984) no. 2, 294-297.

\bibitem{FK} E.V. Ferapontov and K.R. Khusnutdinova, {\it On the integrability of (2+1)-dimensional quasilinear systems}. Comm. Math. Phys. {\bf 248} (2004), no. 1, 187-206.

\bibitem{FK1} E.V. Ferapontov and K.R. Khusnutdinova, {\it The Haantjes tensor and double waves for multi-dimensional systems of hydrodynamic type: a necessary condition for integrability}. Proc. R. Soc. Lond. Ser. A Math. Phys. Eng. Sci. {\bf 462} (2006), no. 2068, 1197-1219.

\bibitem{FKT} E.V. Ferapontov, K.R. Khusnutdinov and S.P. Tsarev, {\it On a class of three-dimensional integrable Lagrangians}. Comm. Math. Phys. {\bf 261} (2006), no. 1, 225-243.

\bibitem{FLS} E.V. Ferapontov, P. Lorenzoni and A. Savoldi,  {\it Hamiltonian operators of Dubrovin-Novikov type in 2D}.  Lett. Math. Phys. {\bf 105} (2015), no. 3, 341-377.

\bibitem{FO} E.V. Ferapontov and A.V. Odesskii, {\it Integrable Lagrangians and modular forms}. J. Geom. Phys. {\bf 60} (2010), no. 6-8, 896-906.

\bibitem{FOS} E.V. Ferapontov, A.V. Odesskii and N.M. Stoilov, {\it Classification of integrable two-component Hamiltonian systems of hydrodynamic type in 2+1 dimensions}. J. Math. Phys. {\bf 52} (2011), no. 7, 073505, 28 pp. 

\bibitem{Grinberg} N.I. Grinberg,  {\it On Poisson brackets of hydrodynamic type with a degenerate metric}. Russian Math. Surveys {\bf 40} (1985), no.4, 231-244.

\bibitem{Mokhov1} O.I.  Mokhov,  \emph{Dubrovin-Novikov type Poisson brackets (DN-brackets)}. Funct. Anal. Appl. {\bf 22} (1988), no. 4, 336-338.

\bibitem{Mokhov2} O.I.  Mokhov,  \emph{The classification of nonsingular multidimensional Dubrovin-Novikov brackets}. Funct. Anal. Appl. {\bf 42} (2008), no.1,  33-44.

\bibitem{Mokhov3} O.I. Mokhov, \emph{Symplectic and Poisson structures on loop spaces of smooth manifolds, and integrable systems}. Russian Mathematical Surveys {\bf 53} (1998), no. 3, 515-622.

\bibitem{Sav} A. Savoldi,  {\it On deformations of one-dimensional Poisson structures of hydrodynamic type with degenerate metric}.  arXiv:1410.3361.

\bibitem{SSY} A.F. Sidorov, V.P. Shapeev, and N.N. Yanenko, {\it The method of differential constraints and its applications in gas dynamics}, Nauka Sibirsk. Otdel., Novosibirsk, (1984), 272 pp.

\bibitem{Strachan1} I.A.B. Strachan, {\it Degenerate Frobenius manifolds and the bi-Hamiltonian structure of rational Lax equations}. J. Math. Phys. {\bf 40} (1999), 5058-5079.

\bibitem{Strachan} I.A.B. Strachan, {\it Degenerate bi-Hamiltonian structures of hydrodynamic type}. (Russian) Teoret. Mat. Fiz. {\bf 122} (2000), no. 2, 294--304; translation in Theoret. and Math. Phys. {\bf 122} (2000), no. 2, 247-255.

\bibitem{Tsarev} S.P. Tsarev, \emph{Geometry of Hamiltonian systems of hydrodynamic type. Generalized hodograph method}. Izvestija AN USSR Math. {\bf 54} (1990), no. 5, 1048-1068.

\end{thebibliography}
\end{document}